\documentclass[pdflatex,sn-mathphys-num]{sn-jnl}

\usepackage{graphicx}%
\usepackage{multirow}%
\usepackage{amsmath,amssymb,amsfonts}%
\usepackage{amsthm}%
\usepackage{mathrsfs}%
\usepackage[title]{appendix}%
\usepackage{xcolor}%
\usepackage{textcomp}%
\usepackage{manyfoot}%
\usepackage{booktabs}%
\usepackage{algorithm}%
\usepackage{algorithmicx}%
\usepackage{algpseudocode}%
\usepackage{listings}%
\usepackage{multicol}

\theoremstyle{thmstyleone}%
\theoremstyle{thmstyletwo}%

\theoremstyle{thmstylethree}%

\begin{document}

\title[Transport at high density]{SMASH: Results from hadronic transport for heavy-ion collisions at high densities}


\author*[1,2, 3,4]{\fnm{Hannah} \sur{Elfner}}\email{h.elfner@gsi.de}

\author[2,3,4,1]{\fnm{Renan} \sur{Góes-Hirayama}}\email{hirayama@itp.uni-frankfurt.de}
\equalcont{These authors contributed equally to this work.}

\affil*[1]{\orgdiv{Department for hot and dense QCD matter}, \orgname{GSI Helmholtz Center for Heavy-in physics}, \orgaddress{\street{Planckstra\ss e 1}, \city{Darmstadt}, \postcode{64291}, \country{Germany}}}

\affil[2]{\orgdiv{Institute for Theoretical Physics}, \orgname{Goethe University}, \orgaddress{\street{Max-von-Laue-Stra\ss e 1}, \city{Frankfurt am Main}, \postcode{60438}, \country{Germany}}}

\affil[3]{\orgdiv{Frankfurt Institute for Advanced Studies}, \orgaddress{\street{Ruth-Moufang-Stra\ss e 1}, \city{Frankfurt am Main}, \postcode{60438}, \country{Germany}}}

\affil[4]{\orgdiv{Helmholtz Research Academy Hesse for FAIR (HFHF)}, \orgaddress{\street{GSI Helmholtz Center,
Campus Frankfurt}, \city{Frankfurt am Main}, \postcode{60438}, \country{Germany}}}


\abstract{This mini-review summarizes the general setup and some highlight results from the hadronic transport approach SMASH (Simulating Many Accelerated Strongly-interacting Hadrons). We start by laying out the software development structures as well as the particle properties and how they are determined by elementary collisions. The different ways to produce light clusters in SMASH, either by coalescence or dynamic multi-particle reactions, are explained. The constraints on nuclear mean fields and the corresponding equation of state from collective flow observables are discussed. In addition, we show how fluctuations associated with a potential critical endpoint survive through the hadronic rescattering stage. Besides hadronic observables, electromagnetic probes offer nice possibilities to study the properties of matter. We present results on collisional broadening of resonances and elliptic flow of dileptons. Last but not least, we review how SMASH can be employed as part of a hybrid approach including a Bayesian analysis for transport coefficients as a function of temperature and chemical potential. We end with an outlook how the hybrid approach has been recently extended to lower collision energies by dynamical fluidization initial conditions. 
}

\keywords{Heavy-ion reactions, Transport theory, Hydrodynamics, Collective behaviour}



\maketitle

\section{Introduction}\label{sec1}

Heavy-ion collisions offer the unique opportunity to study heated and compressed nuclear matter under controlled conditions. In particular collisions of heavy nuclei at low to intermediate beam energies ($\sqrt{s_{\rm NN}}=2.4-7.7$) GeV allow to access the high density regime that is relevant for the dynamics of neutron star mergers. The region of intermediate temperatures and high densities covers a highly interesting part of the phase diagram of quantum chromodynamics, since a critical endpoint and first order transition between the hadron gas and quark-gluon plasma phase are expected \cite{QM25:Rennecke}. The experimental results on net proton fluctuations from the RHIC Beam Energy Scan recently completed by the STAR collaboration \cite{STAR:2025zdq} are not yet conclusive and data is missing in the region accessible for the future FAIR-CBM program. 

To draw any conclusions from final state measurements about the properties of strong-interaction matter, detailed dynamical models are essential. At higher beam energies, the standard model consists of non-equilibrium initial conditions, viscous hydrodynamics and hadronic rescattering dubbed as hybrid approaches \cite{Petersen:2014yqa, Schenke:2020mbo, JETSCAPE:2020mzn, Nijs:2020roc}. In the low energy regime, transport approaches (e.g. JAM, PHSD, UrQMD) are very successful \cite{Nara:1999dz, Cassing:2009vt, Bass:1998ca, Bleicher:1999xi}. The intermediate energy range of interest needs some additional developments concerning mean field potentials, multi-particle interactions or dynamically coupled hybrid approaches with the aim to transition from pure vacuum hadronic physics to a dense system with potentially deconfined degrees of freedom. 

In this mini-review, we give an overview over the SMASH transport approach covering recent results and major developments touching all of the above mentioned items. In Section \ref{sec:smash} the general settings of the model are described, including some comments on sustainable software development, as well as some details on recent updates concerning strangeness production. Section \ref{sec:lightnuclei} covers the production of light nuclei either by coalescence or multi-particle interactions. The following Section \ref{sec:eos} introduces the mean field and its properties including its momentum dependence and results of the Bayesian analysis compared to HADES flow data. In Section \ref{sec:flucs} the fate of potential critical fluctuations in the late hadronic evolution is discussed. The Sections \ref{sec:emprobes} and \ref{sec:resonances} concern the emission of electromagnetic probes and the properties of resonances in a dense medium. The last Section \ref{sec:hybrid} introduces the SMASH-vHLLE hybrid approach and summarizes the main findings of a Bayesian analysis on transport coefficients and the future developments of a dynamical hybrid approach. In the end, there is a summary and an outlook in Section \ref{sec:sum}.

\section{General setup of SMASH and hadron production}\label{sec:smash}

SMASH (Simulating Many Accelerated Strongly-interacting Hadrons) is a hadronic transport approach based on vacuum degrees of freedom \cite{SMASH:2016zqf}. There are more than 150 hadronic species including the corresponding multiplets and antiparticles. Dilepton and photon emission is handled perturbatively. The cross-sections follow experimental data from elementary collisions or are extrapolated using the additive quark model. At low energies, inelastic reactions are treated via resonance excitation and decay, while at high energies a string model is employed \cite{Mohs:2019iee}. By default the geometric collision criterion is applied in a Lorentz covariant manner. Starting from SMASH version 3.1 the total cross-sections are kept fixed, even if the underlying contribution of different channels varies, to ensure a physical cross-section in case of changing resonances and their properties \cite{SanMartin:2023zhv}, as well as a proper treatment of exotic scattering channels. 

As described in detail in \cite{Sciarra:2024gcz} SMASH follows stringent rules for software development. There are descriptive changelogs indicating all changes and physics results for key quantities available for each public release\footnote{\url{https://theory.gsi.de/~smash/analysis_suite/current}}. The codebase is covered by about 100 unit tests to ensure proper running when interfaces change. A wide set of input and output formats is supported corresponding to the major ones used in the field. Even though this is not directly science related, it ensures the reproducibility of scientific results with sophisticated numerical simulations. 

One of the latest improvements of SMASH includes the updated properties of resonances that play a role in strangeness production in elementary collisions. In \cite{Rosenkvist:2025tcl} a genetic algorithm is employed to assess in a systematic way which properties lead to the best description of exclusive cross-section data. Some of the particle properties and branching ratios in SMASH have been adjusted accordingly. The actual list of degrees of freedom and their properties are available in two human readable text files\footnote{\url{https://github.com/smash-transport/smash/blob/main/input/particles.txt} and \url{https://github.com/smash-transport/smash/blob/main/input/decaymodes.txt}}.

\subsection{Production of clusters and light nuclei} 
\label{sec:lightnuclei}

The production of light clusters and nuclei is important at low beam energies, where a large fraction of the nucleons is actually bound in light nuclei and does not exit the interaction region as free protons or neutrons. It is of interest how those nuclei are formed, but also if and how they participate in the collective behaviour. At intermediate energies, ratios of light nuclei yields might also be sensitive to baryon number fluctuations resulting from a critical endpoint or a first order phase transition. 

SMASH incorporates three different ways to produce light nuclei. There is a coalescence mechanism \cite{Mohs:2020awg}, that identifies clusters in the final state particles based on distance in coordinate and momentum space. For that purpose the particles have to be propagated backwards to the time of last interaction. This prescription does not work with many testparticles, therefore when combined with mean fields, one has to use parallel ensembles. The other two ways produce light nuclei dynamically: either only deuterons are generated with a fictitious $d^\prime$ resonance as intermediate state \cite{Oliinychenko:2020znl,Oliinychenko:2018ugs}, or several multiparticle interactions have been implemented that lead to the production and destruction of deuterons, tritons, Helium-3 and hypertritons \cite{Staudenmaier:2021lrg}. Multiparticle interactions rely on stochastic rates, so a statistical collision criterion based on testparticles in cells is used.

\begin{figure}[h]
\centering
\includegraphics[width=0.95\textwidth]{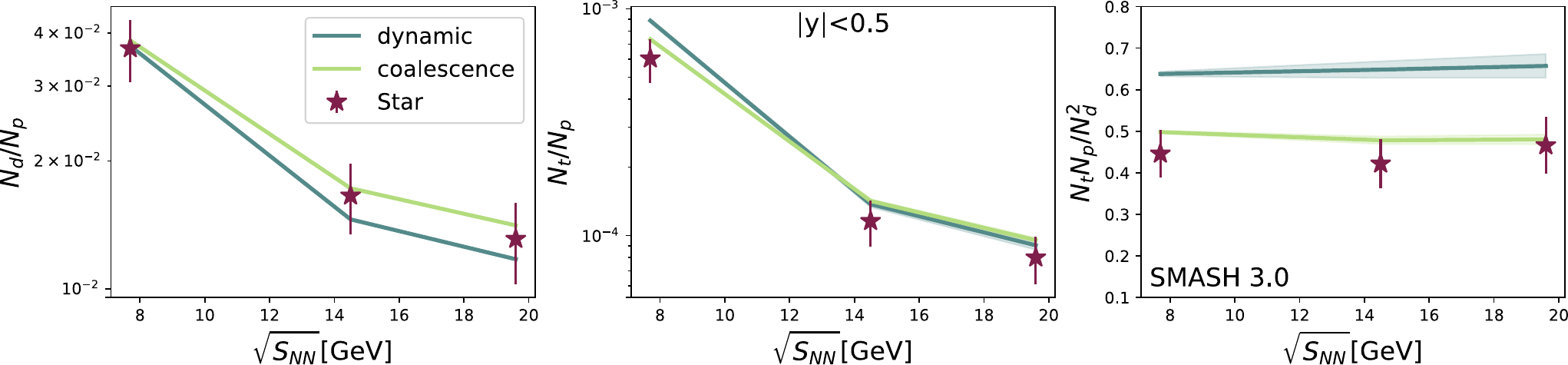}
\caption{Collision energy dependence of particle ratios at mid-rapidity in central AuAu collisions. Data points are taken from \cite{STAR:2022hbp}. [Figure from \cite{Ege:2024vls}]}\label{fig:nuclei}
\end{figure}

As an example, in Fig. \ref{fig:nuclei}  the deuteron and triton production is compared to the proton production in AuAu collisions at three different beam energies in the beam energy scan range from RHIC. The calculated yield ratios within a hybrid approach including light nuclei production either dynamically or by coalescence agree rather well with the experimental measurements. The more detailed results in \cite{Ege:2024vls} indicate that the hadronic rescattering dynamics are crucial to reproduce the transverse momentum spectra of light nuclei. The production of light nuclei has to be also taken into account when looking at the equation of state and comparing to collective flow data, as pursued in the next Section. 

\section{Equation of state of nuclear matter}
\label{sec:eos}

Constraining the equation of state of nuclear matter at high densities with experimental data from heavy-ion collisions is one of the major goals of the community. If the symmetry energy is added the aim is to extrapolate to pure neutron matter such that the insights can be used for the understanding of neutron star mergers in astrophysics. There has been a long standing effort in applying transport approaches with mean fields to obtain such constraints including a systematic evaluation of uncertainties in different models \cite{TMEP:2022xjg, LeFevre:2015paj}. SMASH is following the BUU type implementation, where the potentials are evaluated as a function of local densities evaluated from many test-particles in different ensembles. At this point, there is a Skyrme potential for the nuclear interaction, a symmetry potential and the electromagnetic potential, all in a non-relativistic fashion and therefore only valid in the low beam energy range. The most recent addition is the momentum dependence of the mean field which is described in detail in \cite{Mohs:2024gyc}. 

\begin{figure}[h]
\centering
\includegraphics[width=0.4\textwidth]{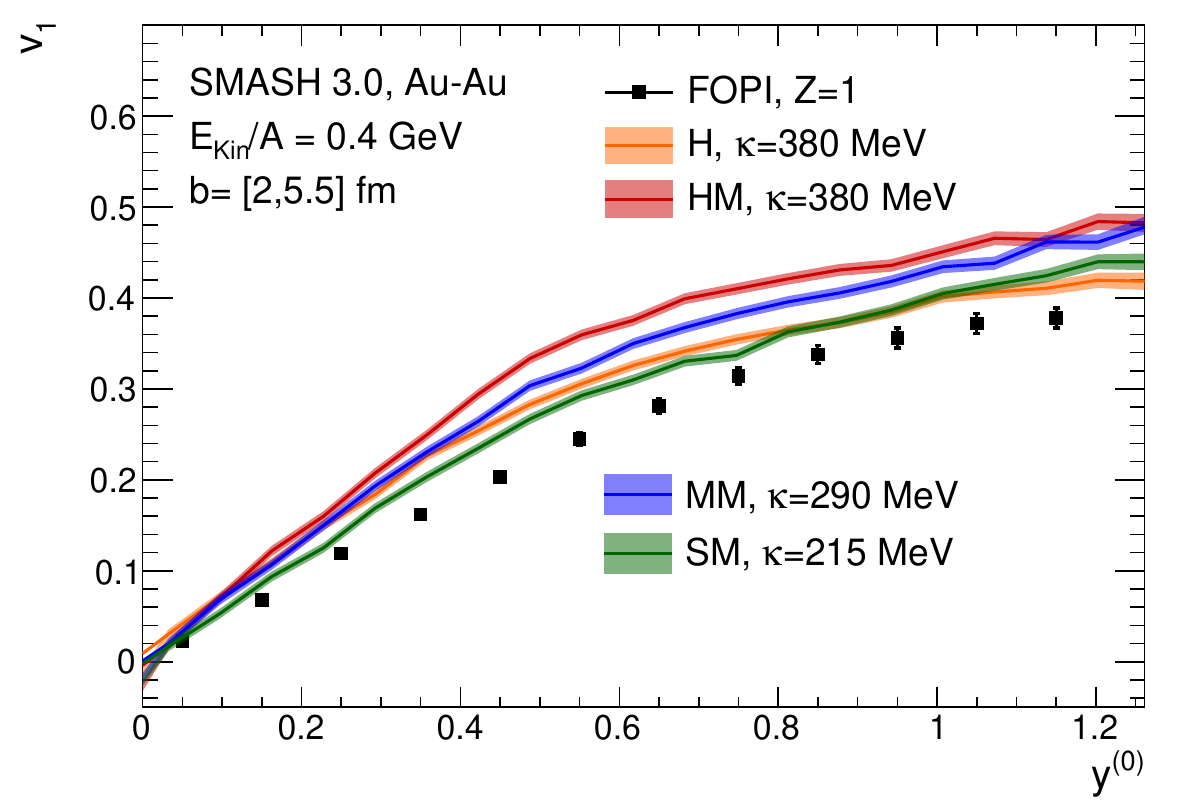}
\includegraphics[width=0.4\textwidth]{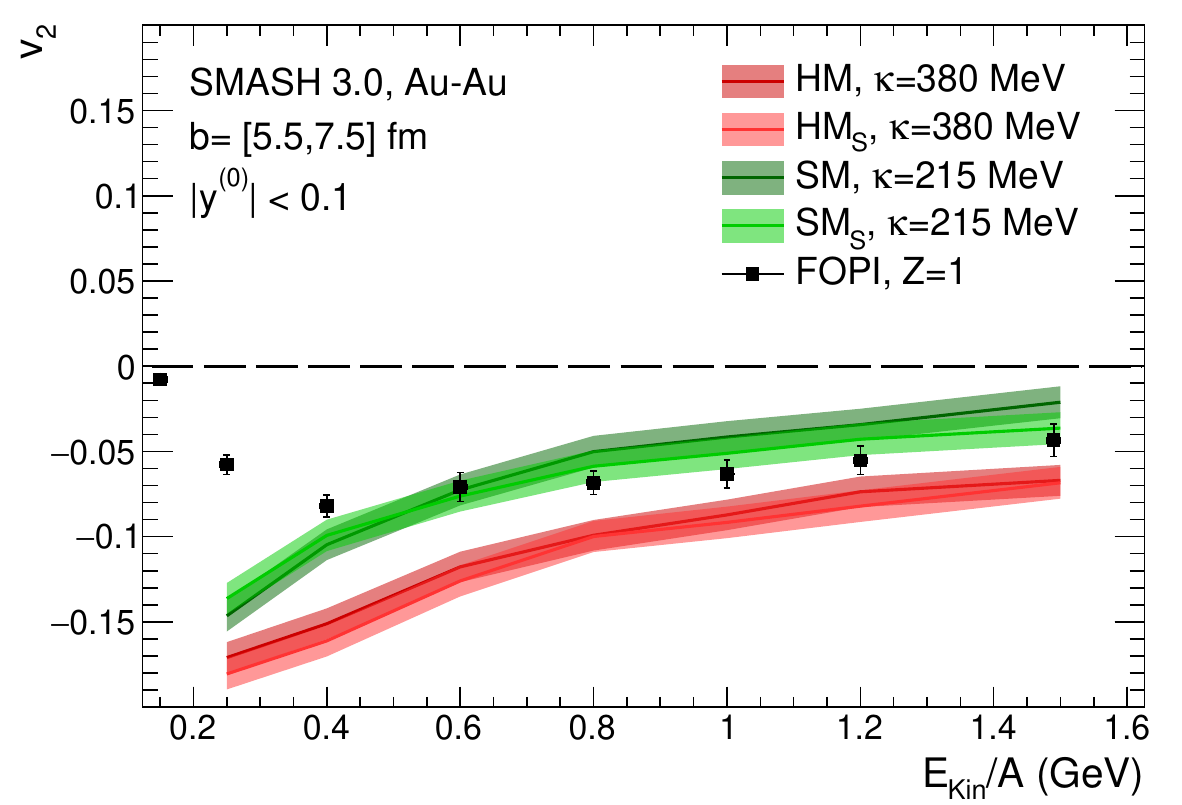}
\caption{Directed and elliptic flow in Au-Au collisions calculated with different settings for the mean field potential compared to FOPI experimental data. [Figures from \cite{Tarasovicova:2024isp}]}\label{fig:fopiflow}
\end{figure}

Figure \ref{fig:fopiflow} shows a comparison of SMASH calculations to directed and elliptic flow for Z=1 particles measured by the FOPI collaboration with different settings for the mean field potential \cite{Tarasovicova:2024isp}. On the left, the directed flow as a function of rapidity is shown in AuAu collisions at $E_{\rm lab}=0.4A$ GeV. The soft momentum dependent potential leads to the best agreement with the data, consistently with prior findings in the literature. On the right hand side, the excitation function of elliptic flow is shown and one can observe that at higher beam energies the hard momentum dependent equation of state is also in decent agreement with the data.

\begin{figure}[h]
\centering
\includegraphics[width=0.4\textwidth]{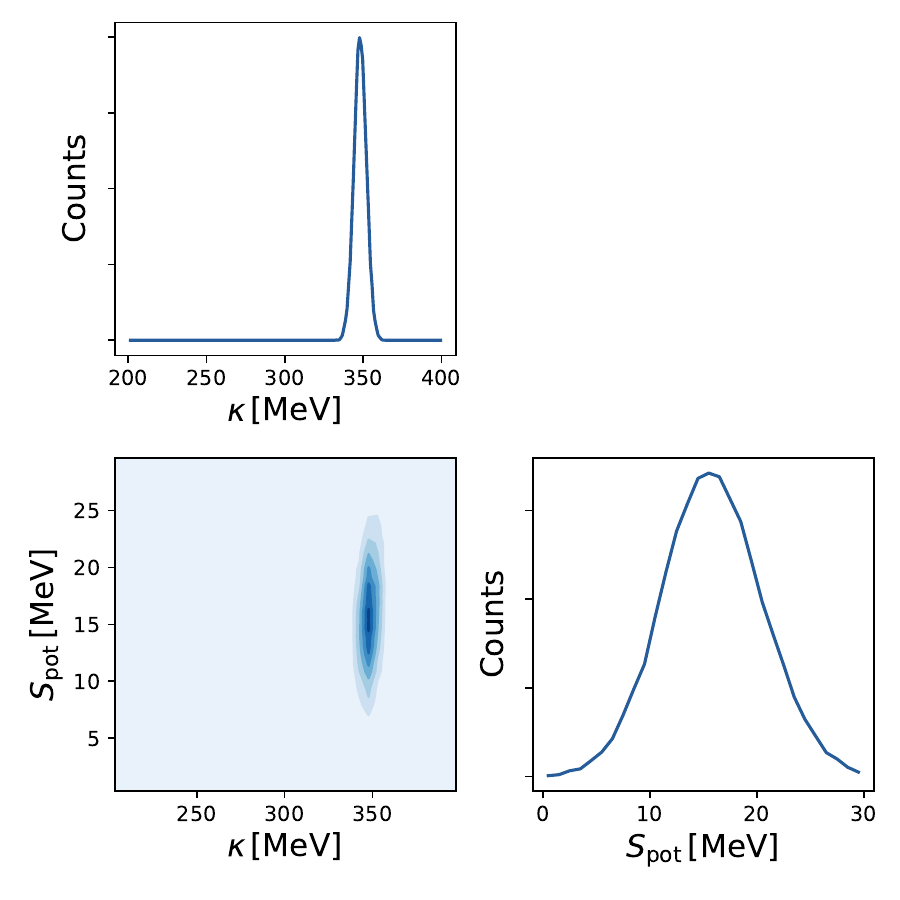}
\includegraphics[width=0.5\textwidth]{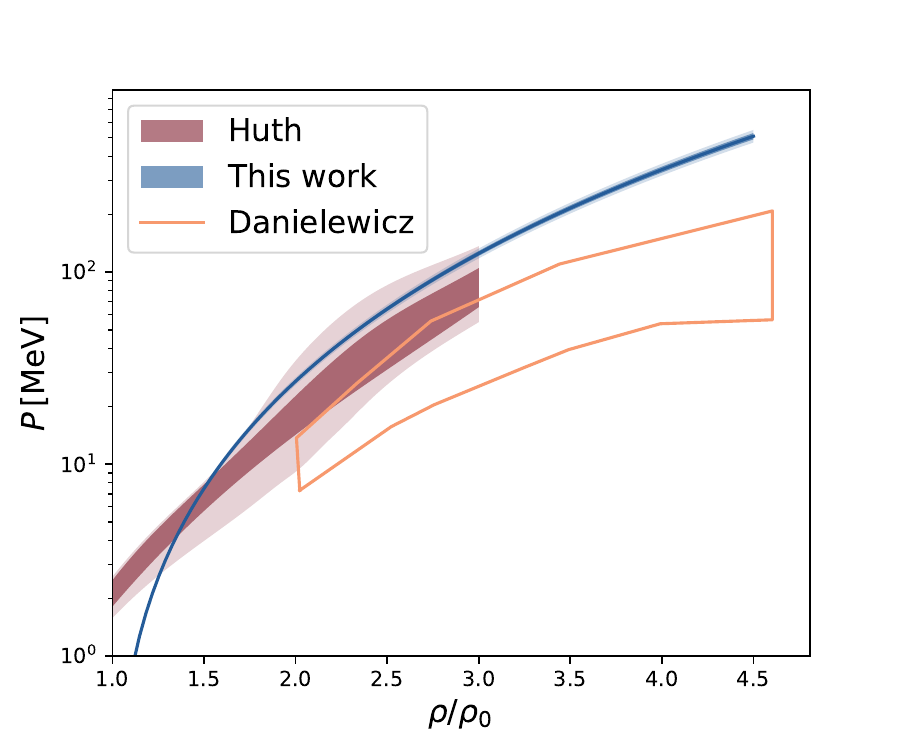}
\caption{Left: Results for the posterior distribution of the incompressibility $\kappa$ and the symmetry energy coefficient from a Bayesian analysis comparing SMASH calculations to HADES collective flow data. Right: Comparison of our results to other constraints in the literature for the equation of state of nuclear matter.  [Figures from \cite{Mohs:2024gyc}]}\label{fig:hadesflow}
\end{figure}

For AuAu collisions at $E_{\rm lab}=1.23A$ GeV the HADES collaboration has taken a large dataset and provided detailed differential measurements of anisotropic flow coefficients. In \cite{Mohs:2024gyc}, we have performed a Bayesian analysis for the symmetry energy parameter and the incompressibility of the equation of state. The momentum dependence has been fixed to match the optical potential. The directed and elliptic flow for protons and deuterons are considered in the intermediate transverse momentum range $1.0<p_\perp < 1.5\ \mathrm{GeV}$, which is less affected by uncertainties due to light nuclei production. The results as shown in Fig. \ref{fig:hadesflow} are consistent with prior findings of other groups including constraints from astrophysics and chiral effective field theory \cite{Huth:2021bsp}. The unexpectedly high incompressibility points to the necessity of considering a more involved density dependence of the potential. 

\subsection{Fate of critical fluctuations}
\label{sec:flucs}

One important landmark in the QCD phase diagram is the critical endpoint, which would imply a first order transition at high densities and lower temperatures. Fluctuations of net baryon number are suggested to be the main signature of critical behaviour. Obvioulsy there are many caveats in mapping equilibrium theory calculations to experimental results from highly dynamical heavy-ion reactions \cite{Bzdak:2019pkr} . 

\begin{figure}[h]
\centering
\includegraphics[width=0.4\textwidth]{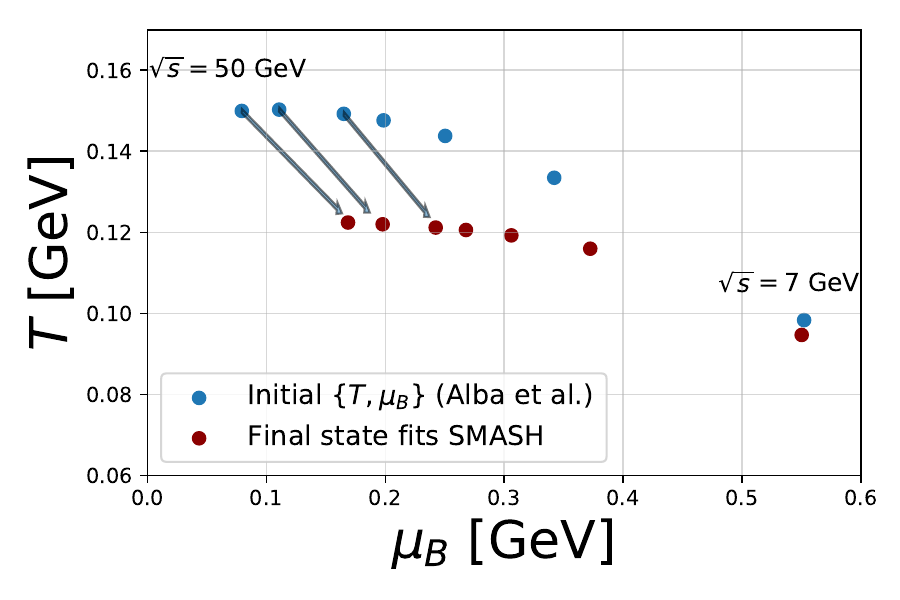}
\caption{Points in the phase diagram that are used as starting points for our calculations as well as the final state fits for $T$ and $\mu_B$. [Figure from \cite{Hammelmann:2023aza}]}\label{fig:trajectories}
\end{figure}

In \cite{Hammelmann:2023aza} we have considered the effect of hadronic rescattering on higher moments of the net proton distribution that were recently measured in the beam energy scan by STAR \cite{STAR:2025zdq}. In Fig. \ref{fig:trajectories} the start and end points of our hadronic evolution are indicated in the temperature-chemical potential plane. The initial values correspond to chemical freeze-out fits to the measured yields. A hadron gas corresponding to those values has been enhanced with the information on possible critical fluctuations from an Ising model and the subsequent hadronic evolution of an expanding sphere has been calculated. 

\begin{figure}[h]
\centering
\includegraphics[width=0.77\textwidth]{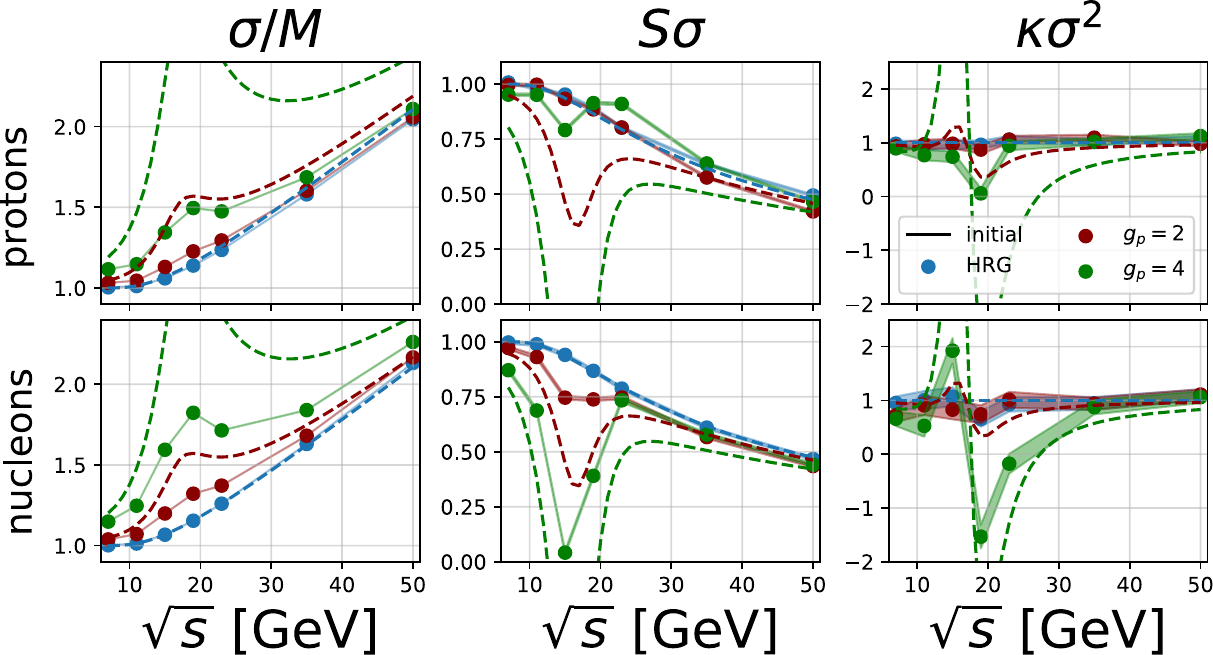}
\caption{Comparison of initial and final state values for the cumulants of the net proton and net baryon distributions for different values of the coupling constant to the critical field. [Figure from \cite{Hammelmann:2023aza}]}\label{fig:cumulants}
\end{figure}

Fig. \ref{fig:cumulants} shows the effect of the hadronic evolution on the different cumulants of the net proton distribution. Note that this is a pessimistic scenario, since the correlations are not propagated in the SMASH hadronic transport approach. Nevertheless, for a larger but still reasonable coupling of the critical mode, the signals survive the rescattering evolution. 

\section{Electromagnetic probes}
\label{sec:emprobes}

\begin{figure}[h]
\centering
\includegraphics[width=0.4\textwidth]{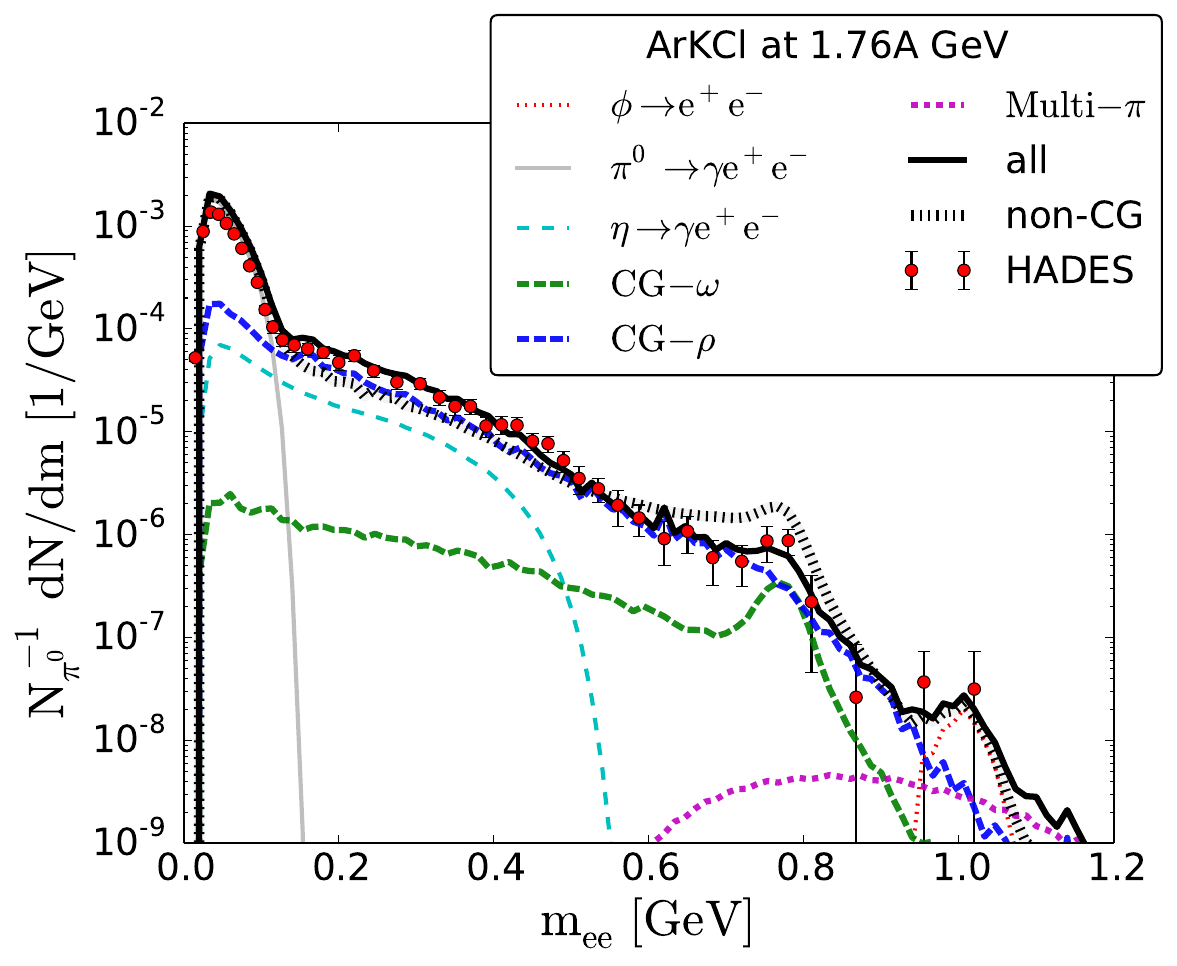}
\includegraphics[width=0.4\textwidth]{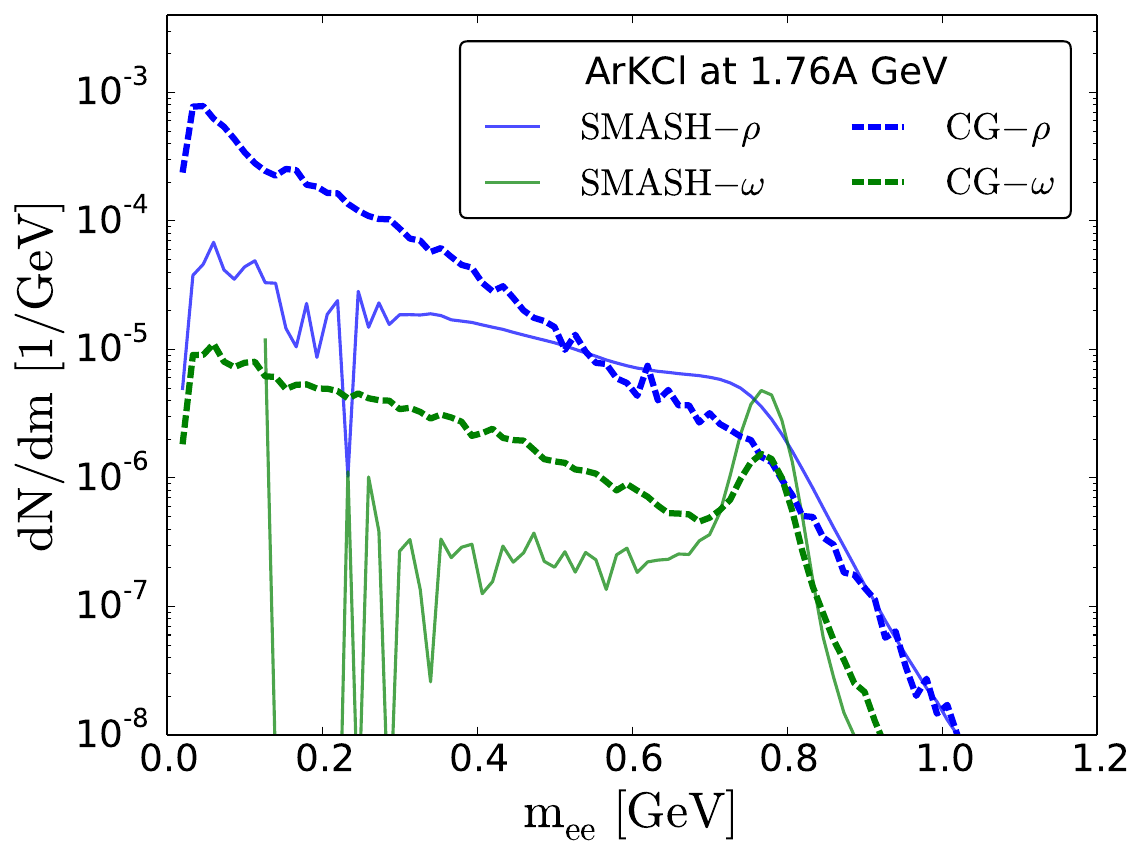}
\caption{Left: Invariant mass distribution of dileptons in Ar+KCl collisions compared to HADES experimental data. The full line depicts the result including a full medium modified spectral function in a coarse-grained simulation, while the dashed line represents a pure transport calculation. Right: Comparison of the contributions of the vector mesons within both calculations. [Figures from \cite{Staudenmaier:2017vtq}]}\label{fig:dileptons}
\end{figure}

Penetrating probes such as dileptons are interesting observables, since they are emitted from all stages of the reaction and might therefore track a potential phase transition nicely \cite{Seck:2020qbx}. In the beam energy range of interest, CBM at FAIR will provide the first results for dilepton measurements. At lower energies the HADES collaboration has measured dileptons in different systems and a comprehensive theoretical analysis of this data has been performed in \cite{Staudenmaier:2017vtq}. Fig. \ref{fig:dileptons} shows the invariant mass spectrum of dileptons in Ar+KCl collisions at $E_{\rm lab} = 1.76A $ GeV. Even at this energy, the medium is already so hot and dense that the best description of the experimental data requires medium-modified spectral functions. Within SMASH, there is an apples-to-apples comparison (Fig. \ref{fig:dileptons} right) for the $\rho$ and $\omega$ mesons between the vacuum Breit-Wigner spectral functions and the ones employed in a coarse-grained calculation including the medium effects. The peaks at the pole mass are very diluted and the distribution is much broader in general. 

\begin{figure}[h]
\centering
\includegraphics[width=0.42\textwidth]{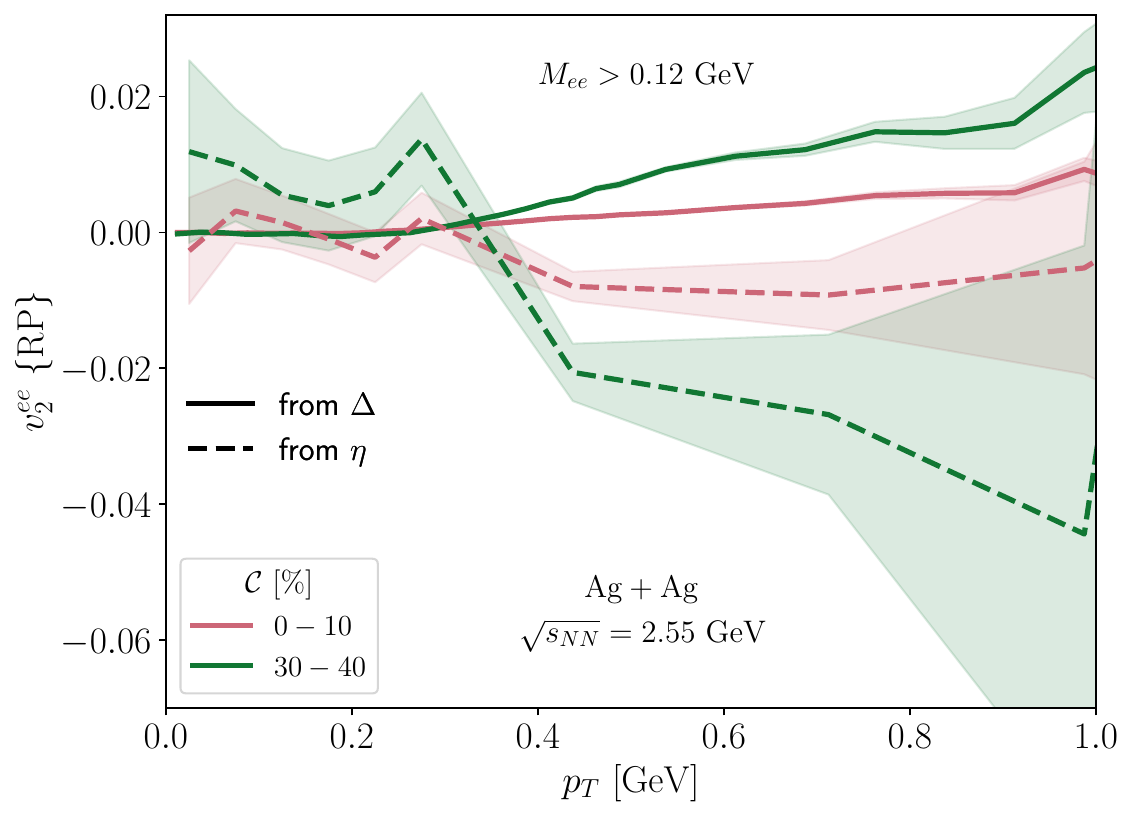}\hspace{0.02\linewidth}
\includegraphics[width=0.42\textwidth]{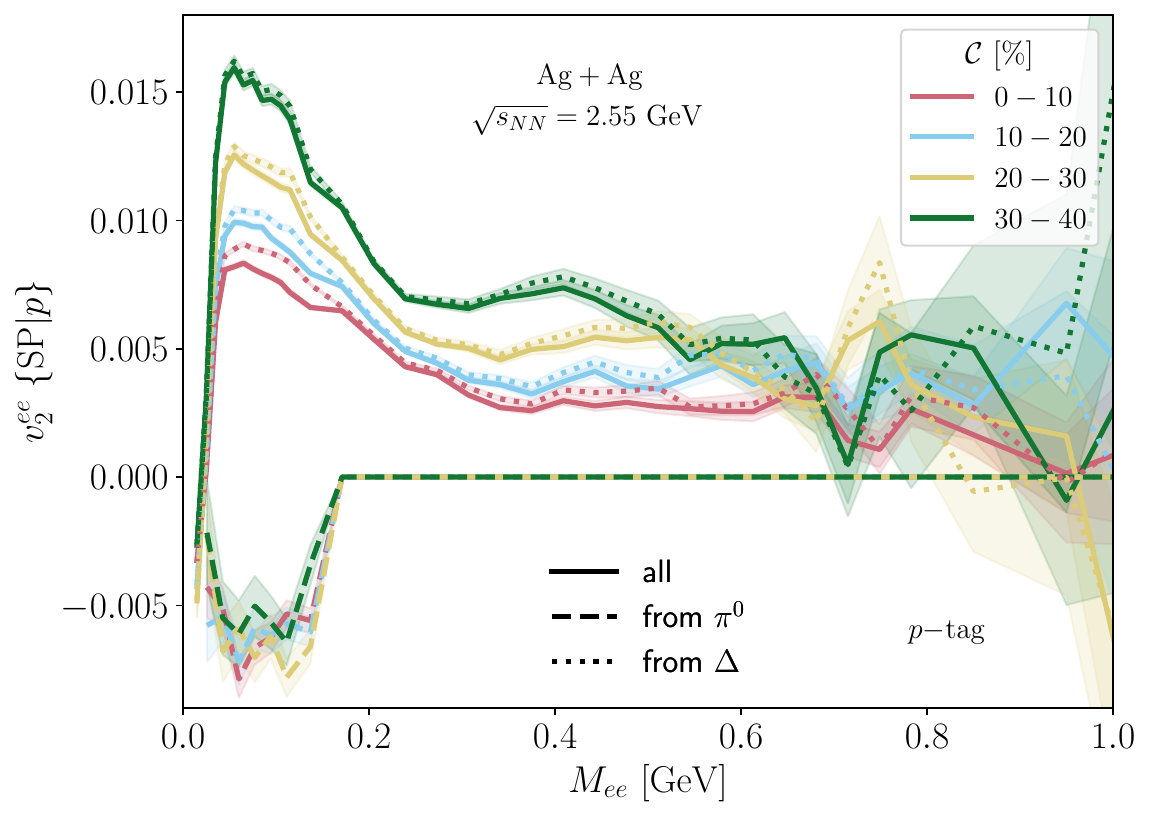}
\caption{Elliptic flow of dileptons in Ag+Ag collisions at $\sqrt{s_{NN}}=2.55\ \mathrm{GeV}$ and different centrality classes. Left: Contribution of the $\Delta$ baryon and $\eta$ meson to the reaction plane $v_2$ of dileptons above the pion threshold. Right: Alignment of the total flow with the $\Delta$ contribution in the proposed $p$-tagged Scalar Product method. [Figures from  \cite{Goes-Hirayama:2024aqz}]}\label{fig:dilepton_flow}
\end{figure}

Dileptons can also be used to access detailed information about the composition of the system, since each resonance contributes to the dilepton emission at different stages of the expansion. For example, as shown in Fig. \ref{fig:dilepton_flow}, the $\Delta\to N e^+e^-$ and $\eta\to \gamma e^+e^-$ decays have opposing contributions to the total flow \cite{Goes-Hirayama:2024aqz}. The small and positive initial thermal flow was also seen in \cite{Reichert:2023eev}, and the negative final state flow is consistent with the squeeze-out picture. The overall reaction plane flow suffers from cancelling effects, leading to the $v_2^{ee}$ consistent with 0 measured in preliminary HADES results \cite{Schild:2024ywo}.

Accessing resonance dynamics from final state measurements can be achieved by a careful construction of the reference plane. For that, we proposed the Tagged Scalar Product method in \cite{Goes-Hirayama:2024aqz}, which correlates the total dilepton emission with a specific hadron species. The right plot of fig. \ref{fig:dilepton_flow} shows that, when the $p$-tagged method is used, the dilepton flow is very closely aligned with the flow from $\Delta$ radiation, because protons are only produced alongside dileptons in the $\Delta$ decay. This method therefore opens a window to directly study the dynamics of some resonances via dileptons.

\subsection{Collisional broadening of resonances}
\label{sec:resonances}

While the dilepton spectra from SMASH shown as non-CG in fig. \ref{fig:dileptons} uses vacuum spectral functions, without \emph{thermal} modifications, there are modifications due to collisional broadening within the hadronic medium: if the $\rho$ and $\omega$ are absorbed, the resulting dilepton yield is suppressed. This is true for any resonance that has a non-zero inelastic cross section. Defining the effective width $\Gamma^\mathrm{eff}=\langle\tau\rangle^{-1}$, with $\tau$ being a resonance lifetime, we can extract the corresponding broadened spectral function as a Breit-Wigner distribution that uses $\Gamma^\mathrm{eff}$. 

\begin{figure}[h]
\centering
\includegraphics[width=0.4\textwidth]{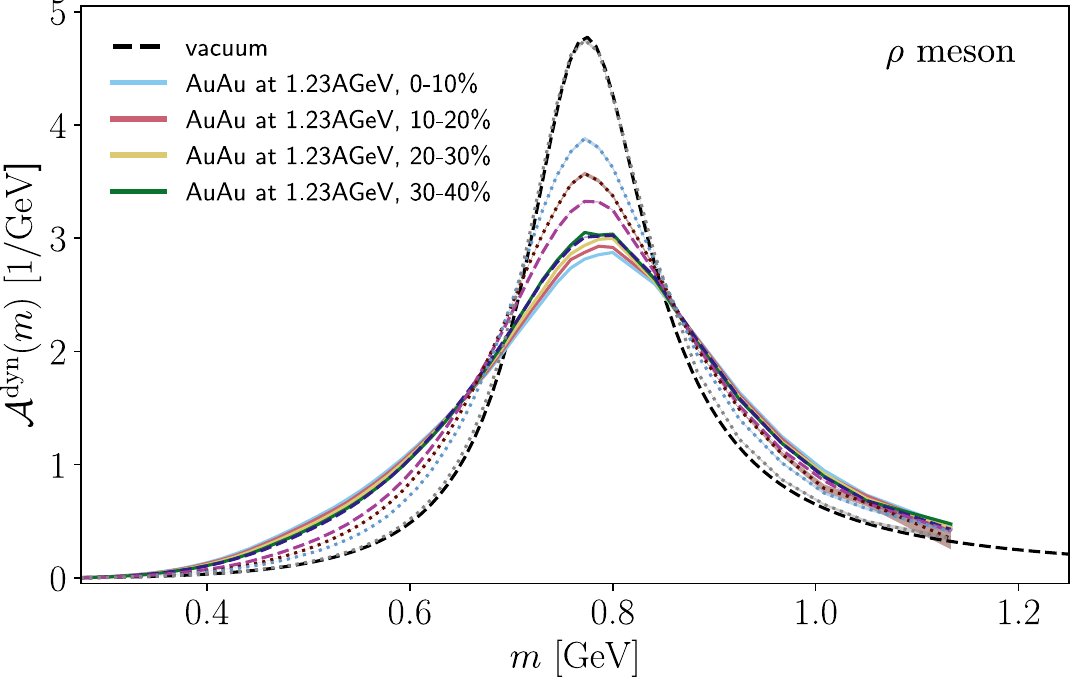}\hspace{0.02\linewidth}
\includegraphics[width=0.415\textwidth]{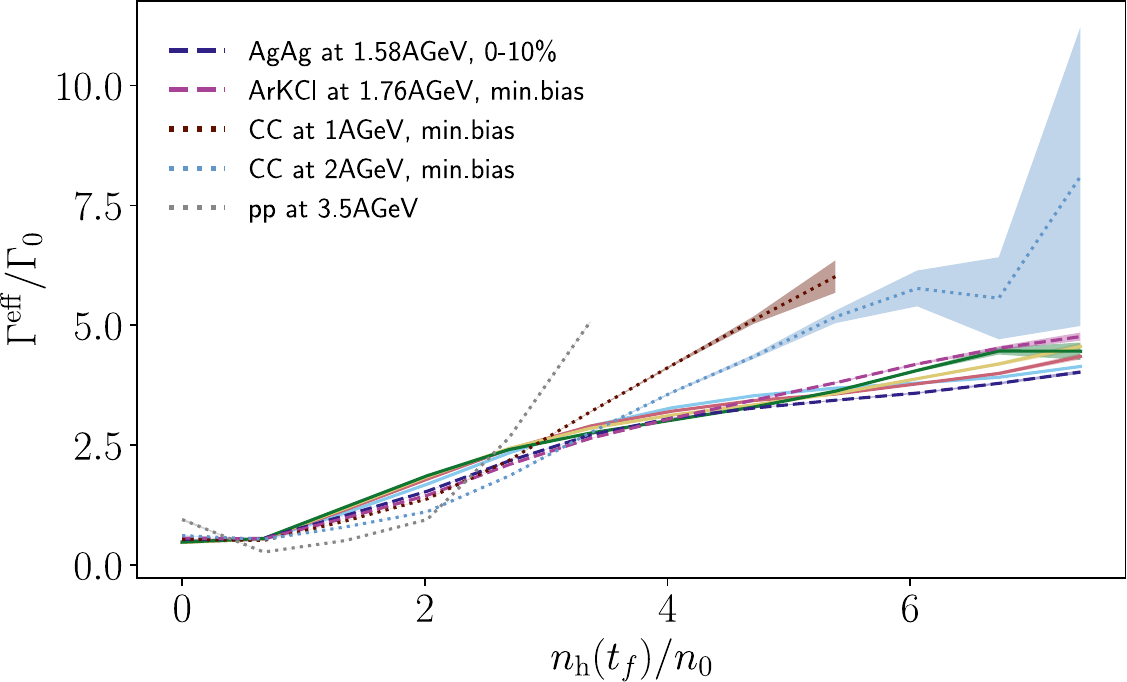}
\caption{Left: Spectral function of the $\rho$ meson in vacuum and different collision systems. Right: Universal scaling of the effective width with the local hadron density. [Figures from \cite{Hirayama:2022rur}]}\label{fig:coll_broad}
\end{figure}

The left plot of fig. \ref{fig:coll_broad} quantifies this for the $\rho$ in several collision systems corresponding to the ones performed by the HADES experiment. A clear hierarchy appears where collisional broadening increases in larger systems, a consequence of the quasi-universal density dependence of the effective width shown in the right plot, where all systems fall on or are close to the same curve. This is consistent with assumptions in in-medium transport models, such as GiBUU and PHSD \cite{Buss:2011mx,Bratkovskaya:2007jk}, that use a density dependent width corresponding to collisional broadening for the resonances. 

\section{SMASH-vHLLE hybrid approach}
\label{sec:hybrid}
For higher beam energies, SMASH has been coupled to the 3+1 dimensional viscous hydrodynamics code vHLLE. The hybrid approach consists of initial conditions from SMASH, where the nuclei are initialized according to Woods-Saxon profiles and the subsequent microscopic dynamics is calculated until a certain eigentime $\tau_0$ is reached. After the hot and dense evolution within hydrodynamics hadrons are sampled on the Cooper-Frye hypersurface at a constant energy density and the subsequent hadronic evolution is again calculated within SMASH \cite{Schafer:2021csj}. The whole framework for the hybrid approach has been incorporated in a bash handler that is publicly available \cite{sciarra_2025_15880337}. 

\begin{figure}[h]
\centering
\includegraphics[width=0.43\textwidth]{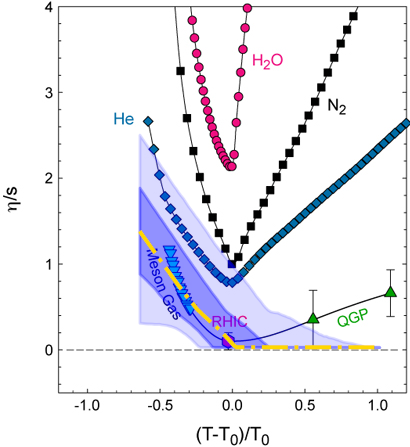}
\includegraphics[width=0.4\textwidth]{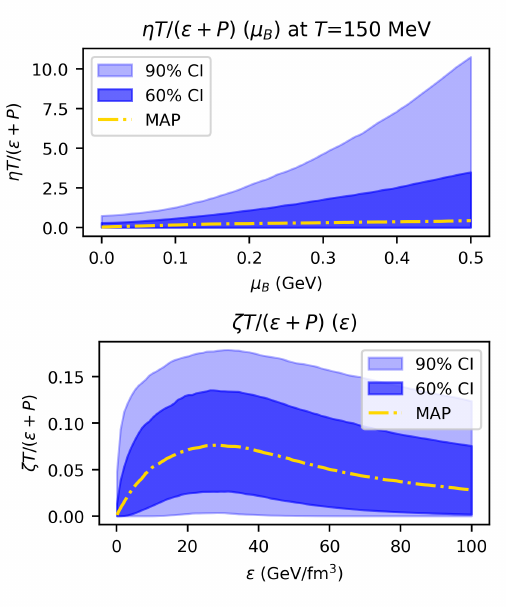}
\caption{Shear and bulk viscosity over entropy density as a function of temperature (and net baryon chemical potential) extracted from a Bayesian analysis comparing the SMASH-vHLLE hybrid approach to RHIC beam energy scan bulk observables. [Figures from \cite{Gotz:2025wnv}]
}\label{fig:transportcoeff}
\end{figure}

While \cite{Gotz:2022naz} contains a first glimpse at how important an explicit density dependence of transport coefficients is, in \cite{Gotz:2025wnv} we have performed a full Bayesian analysis with respect to a large set of experimental data for bulk observables from the RHIC beam energy scan. From this, we extract the temperature and net baryon chemical potential dependence of shear and bulk viscosity. The main results for the posterior distributions are shown in Fig. \ref{fig:transportcoeff}. Surprisingly, very low values of shear viscosity are required at higher temperatures while the baryon chemical potential dependence is not much constrained. The temperature dependence of the bulk viscosity is in accordance with prior findings of other analysis \cite{Jahan:2024wpj}.

To extend the hybrid approach down to lower beam energies, a more dynamical initial state is crucial, since the collision happens over a longer time scale due to the smaller Lorentz contraction, meaning that some parts of the nuclei will form a fluid before others, and some will remain entirely out of equilibrium. 

\begin{figure}[h]
\centering
\includegraphics[width=0.5\textwidth]{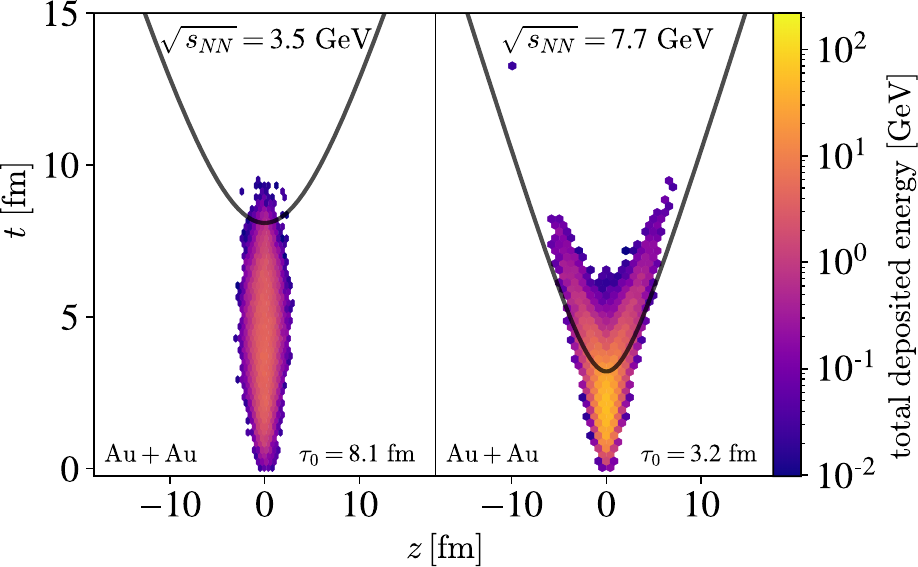}\hspace{0.02\textwidth}
\includegraphics[width=0.41\textwidth]{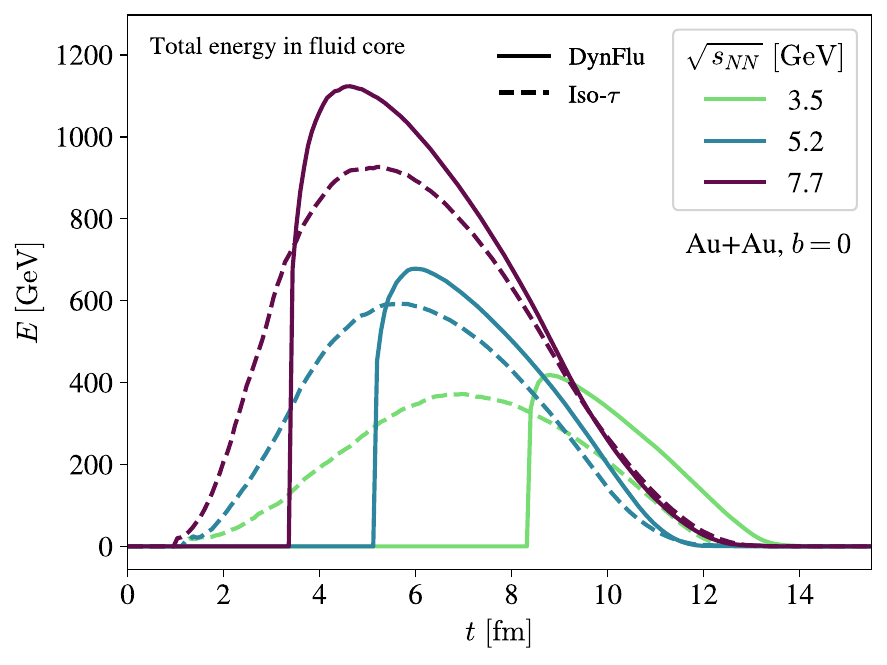}
\caption{Deposition of energy in the hybrid approach for head-on Au+Au collisions at low beam energies. Left: Spacetime dependence of the core-corona separation. Right: Evolution of the total energy contained in the fluid. [Figures from \cite{Goes-Hirayama:2025nls}]}\label{fig:dynflu}
\end{figure}

In \cite{Goes-Hirayama:2025nls} we introduce a core-corona separation based on the local energy density, where SMASH hadrons are dynamically fluidized, with their energy, momentum, and conserved charges used as sources for the hydrodynamic evolution. This is similar to prior work in \cite{Akamatsu:2018olk}. Figure \ref{fig:dynflu} displays the energy deposition in this procedure. The left plots represent where it happens in SMASH, compared to the eigentime $\tau_0$. The right plot shows how the total energy evolves in the fluid for both initialization models. The dynamically initialized fluid is smoother and therefore longer-lived, in particular at the lowest beam energies.

\section{Summary and Outlook}
\label{sec:sum}

To summarize, SMASH is a multi-purpose hadronic transport approach under active development. The microscopic hadronic dynamics is well suited to understand low energy heavy-ion collisions and the hadronic rescattering at high beam energies. In the intermediate energy range hybrid approaches are very successful. We suggest a dynamical initial state based on SMASH to eventually study the phase transition dynamics and its consequences on hadronic and electromagnetic observables, in particular in the energy range that is of interest for the future CBM program at FAIR. 

\backmatter

\bmhead{Acknowledgements}

The data sets needed to reproduce the figures in this article are available via the respective references of the original publications or by contacting the authors. 
Computational resources have been provided by the GreenCube at GSI and the Center
for Scientific Computing (CSC) at the Goethe-University of Frankfurt. 
Funded by the Deutsche Forschungsgemeinschaft (DFG, German Research Foundation) – Project number 315477589 – TRR 211.


\bibliography{SMASH-mini-review-refs}

\end{document}